\documentclass[aps,prl,twocolumn,showpacs]{revtex4-1}
\usepackage{color, graphicx}
\usepackage{amsmath, amssymb}
\usepackage{ulem}

\begin{document}

\title{Multipartite Entanglement Witnesses}
\author{J. Sperling}\email{jan.sperling2@uni-rostock.de}\affiliation{Arbeitsgruppe Theoretische Quantenoptik, Institut f\"ur Physik, Universit\"at Rostock, D-18051 Rostock, Germany}
\author{W. Vogel}\affiliation{Arbeitsgruppe Theoretische Quantenoptik, Institut f\"ur Physik, Universit\"at Rostock, D-18051 Rostock, Germany}
\date{\today}

\begin{abstract}
	We derive a set of algebraic equations, the so-called multipartite separability eigenvalue equations.
	Based on their solutions, we introduce a universal method for the construction of multipartite entanglement witnesses.
	We witness multipartite entanglement of $10^3$ coupled quantum oscillators, by solving our basic equations analytically.
	This clearly demonstrates the feasibility of our method for studying ultrahigh orders of multipartite entanglement in complex quantum systems.
\end{abstract}
\maketitle

	Entanglement represents a fundamental quantum correlation between compound quantum systems.
	Since the early days of quantum physics this property has been used to illustrate the surprising consequences of the quantum description of nature~\cite{epr,schroed}.
	Moreover, entanglement plays a fundamental role in various applications and protocols in quantum information science~\cite{nielsenchuang,horodecki,guehne}.

	In multipartite systems a separable state is a statistical mixture of product states~\cite{werner}.
	A quantum state  is entangled, whenever it cannot be represented in this form.
	Various forms of multipartite entanglement are known~\cite{NN1,NN2,NN6,NN9}.
	The most prominent and nonequivalent forms of entangled multipartite quantum states are the GHZ-state~\cite{ghz} and the W-state~\cite{wstate}, which have been generalized to so-called cluster and graph states~\cite{raussendorf:1,raussendorf:2}.
	Another classification is given in terms of partial and full (or genuine) multipartite entanglement, for an introduction see e.g.~\cite{horodecki,guehne}.
	Beyond finite dimensional systems, multipartite quantum entanglement in continuous variable systems turns out to be a cumbersome problem. 
	Even in the case of Gaussian states, there exist multipartite entangled states, which cannot be distilled~\cite{wernerwolf}.

	High orders of multipartite entanglement are of great interest, for example, in quantum metrology.
	Multipartite entanglement has been shown to be essential to reach the maximal sensitivity in metrological tasks~\cite{GZNEO2010}.
	In this context, the quantum Fisher information has been used to characterize the entanglement~\cite{KSWWHPS2012,HLKSWWPS2012,T12}.

	The detection of entanglement is typically done via the construction of proper entanglement witnesses~\cite{witness:1,witness:2,witness:3}, being equivalent to the method of positive, but not completely positive maps.
	A witness is an observable, which is non-negative for separable states, and it can have a negative expectation value for entangled states.
	For different kinds of entanglement, different types of witnesses have been considered: bipartite witnesses~\cite{witness:2,SV09a}; Schmidt number witnesses~\cite{bruss,SV11b}; and multipartite witnesses for partial and genuine entanglement~\cite{NN7,NN5,NN8,LF03}.
	A systematic approach for witnessing entanglement in complex quantum systems is missing yet.

	Recently, we considered the construction of bipartite entanglement witnesses with the so-called separability eigenvalue equations~\cite{SV09a}.
	We have shown that the same equations need to be solved to obtain entanglement quasiprobabilities, which are nonpositive distributions if and only if the corresponding quantum state is entangled~\cite{SV09b}.
	Moreover, we have shown that the Schmidt number witnesses can be obtained by solving the related Schmidt number eigenvalue problem~\cite{SV11b}.

	In the present Letter we derive a set of algebraic equations, which yield the construction of arbitrary multipartite entanglement witnesses.
	For these so-called multipartite separability eigenvalue equations, we will study some fundamental properties, which uncover the structure of multipartite entanglement.
	Examples are given to witness partial and full entanglement in multipartite composed systems, for pure and mixed quantum states in discrete and continuous variable systems.

	In the following, we consider a composed Hilbert space $\mathcal H=\mathcal H_1\otimes\dots\otimes\mathcal H_N$.
	It has been shown that, without loss of generality, we could assume that the individual subsystems are finite dimensional ones~\cite{SV09c}.
	Let us consider a partition $\mathcal I_1\dots\mathcal I_K$ of the index set $\mathcal I=\{1,\dots,N\}$.
	A quantum state $\hat\sigma$ is separable for the given partition, if it can be written as a classical mixture of product states~\cite{werner}:
	\begin{align}\label{Eq:Sep}
		\hat\sigma=\int_{\mathcal S_{\mathcal I_1:\dots:\mathcal I_K}}\hspace*{-4ex} {\rm d}P(a_1,\dots, a_K)\,|a_1,\dots,a_K\rangle\langle a_1,\dots,a_K|,
	\end{align}
	with $P$ being a classical probability distribution and $\mathcal S_{\mathcal I_1:\dots:\mathcal I_K}$ being the set of pure and normalized separable states.
	If a quantum state $\hat\varrho$ cannot be written in the form of Eq.~\eqref{Eq:Sep}, it is referred to as multipartite entangled.

	A multipartite entanglement witness for the given partition is a Hermitian operator $\hat W$, with
	\begin{align}
		\nonumber \langle \hat W\rangle={\rm tr}(\hat\sigma\hat W)\geq0,&\text{ for all $\hat\sigma$ separable,}\\
		\langle \hat W\rangle={\rm tr}(\hat\varrho\hat W)<0,&\text{ for at least one $\hat\varrho$.}\label{Eq:TrueWitness}
	\end{align}
	Based on Refs.~\cite{toth2005,SV09a}, it can readily be shown that any witness can be presented in the form
	\begin{align}
		\hat W=f_{\mathcal I_1:\dots:\mathcal I_K}(\hat L)\hat 1-\hat L,
	\end{align}
	where $\hat L$ is a general Hermitian operator and the function $f_{\mathcal I_1:\dots:\mathcal I_K}(\hat L)$ denotes the maximally attainable expectation value for separable states:
	\begin{align}
		\nonumber f_{\mathcal I_1:\dots:\mathcal I_K}(\hat L)=\sup\{\langle a_1,\dots,a_K|\hat L|a_1,\dots,a_K\rangle\}.
	\end{align}
	The supremum is taken over all $|a_1,\dots,a_K\rangle\in\mathcal S_{\mathcal I_1:\dots:\mathcal I_K}$.

	Hence, we can formulate a necessary and sufficient entanglement criterion being equivalent to the witness criterion:
	A quantum state $\hat\varrho$ is entangled with respect to the partition $\mathcal I_1,\dots\mathcal I_K$, if and only if there exists a Hermitian operator $\hat L$ such that
	\begin{align}\label{Eq:Supremum}
		{\rm tr}(\hat\varrho\hat L)>f_{\mathcal I_1:\dots:\mathcal I_K}(\hat L).
	\end{align}
	This means that the mean value of $\hat L$ exceeds the boundary of mean values for separable states.
	A replacement $\hat L\mapsto-\hat L$ leads to a similar entanglement criterion, but with the greatest lower bound ($\inf$)  instead of the least upper bound ($\sup$):
	\begin{align}\label{Eq:Infimum}
		{\rm tr}(\hat\varrho\hat L)<\inf\{\langle a_1,\dots,a_K|\hat L|a_1,\dots,a_K\rangle\}.
	\end{align}

	For both entanglement criteria, we have to solve the following optimization problem for an observable $\hat L$:
	\begin{align}
		\nonumber &G=\langle a_1,\dots,a_K|\hat L|a_1,\dots,a_K\rangle\to{\rm optimum},\\
		&C=\langle a_1,\dots,a_K|a_1,\dots,a_K\rangle-1\equiv 0,
	\end{align}
	where $G$ represents the function to be optimized, and $C$ is the normalization condition.
	For such an optimization problem, we can apply the method of Lagrangian multipliers.
	In our case, the optimization condition is
	\begin{align}\label{Eq:Lagrange}
		0=\frac{\partial G}{\partial \langle a_j|}-g\frac{\partial C}{\partial \langle a_j|}, \quad\text{ for $j=1,\dots,K$},
	\end{align}
	where $g$ is the Lagrangian multiplier and $0$ is the null vector in the subspace given by the partition $\mathcal I_j$.
	The partial derivatives of $G$ can be computed as:
	\begin{widetext}
		\begin{align}
			\nonumber \frac{\partial G}{\partial \langle a_j|}=&\frac{\partial \langle a_1,\dots,a_N|\hat L|a_1,\dots,a_N\rangle}{\partial \langle a_j|}
			=\frac{\partial\left[{\rm tr}_{\mathcal I_1}\dots {\rm tr}_{\mathcal I_K}\left(\hat L | a_1,\dots,a_K\rangle\langle a_1,\dots,a_K|\right)\right]}{\partial \langle a_j|}\\
			=&{\rm tr}_{\mathcal I_1}\dots {\rm tr}_{\mathcal I_{j-1}}{\rm tr}_{\mathcal I_{j+1}}\dots {\rm tr}_{\mathcal I_K}\left(\hat L \left[|a_1,\dots,a_{j-1}\rangle\langle a_1,\dots,a_{j-1}|\otimes\hat 1_{\mathcal I_j}\otimes |a_{j+1},\dots,a_K\rangle\langle a_{j+1},\dots,a_K|\right]\right) |a_j\rangle\nonumber\\
			=&\hat L_{a_1,\dots,a_{j-1},a_{j+1},\dots,a_K}|a_j\rangle.
		\end{align}
	\end{widetext}
	The case $\hat L=\hat 1$ yields the derivatives of $C$.
	Let us also note that we assumed that the indices of the sets $\mathcal I_j$ are ordered in a form that all elements of $\mathcal I_{j}$ are larger then the elements of $\mathcal I_{j'}$ for $j>j'$.
	This assumption is justified by the fact that one can employ, without loss of generality, a permutation of the Hilbert spaces $\mathcal H_1\dots\mathcal H_N$ to order them in the required form.

	The Euler-Lagrangian optimization condition in Eq.~\eqref{Eq:Lagrange} can be reformulated for all $j=1,\dots,K$ as 
	\begin{align}\label{Eq:Lagrange2}
		0=\hat L_{a_1,\dots,a_{j-1},a_{j+1},\dots,a_K}|a_j\rangle-g |a_j\rangle,
	\end{align}
	where all eigenstates are normalized ones, $\langle a_j|a_j\rangle=1$.
	In addition, we may evaluate the value of $g$.
	We can do this by multiplying Eq.~\eqref{Eq:Lagrange2} with $\langle a_j|$.
	This yields
	\begin{align}
		\nonumber g=&\langle a_j|\hat L_{a_1,\dots,a_{j-1},a_{j+1},\dots,a_K}|a_j\rangle\\
		=&\langle a_1,\dots,a_K|\hat L|a_1,\dots,a_K\rangle=G_{\rm optimum}.
	\end{align}
	Hence, the Lagrangian multiplier corresponds to an optimal expectation value of $\hat L$ for separable states.
	In conclusion of this derivation, we get an algebraic problem whose solutions give all optimal expectation values.
	\paragraph*{Definition: MSEvalue equations.--}
	The equations
	\begin{align*}
		\hat L_{a_1,\dots,a_{j-1},a_{j+1},\dots,a_K}|a_j\rangle=g |a_j\rangle
		\,\text{ for $j=1,\dots,K$}
	\end{align*}
	are defined as the first form of the multipartite separability eigenvalue (MSEvalue) equations.
	The value $g$ is denoted as the MSEvalue of $\hat L$, and the product vector $|a_1,\dots,a_K\rangle$ is the corresponding multipartite separability eigenvector (MSEvector).

	As a final conclusion from this derivation we get
	\begin{align}
		f_{\mathcal I_1:\dots:\mathcal I_K}(\hat L)=\sup\left\{g: \text{ $g$ is MSEvalue of $\hat L$}\right\},
	\end{align}
	and condition~\eqref{Eq:Infimum} is given by the infimum of all MSEvalues.
	This means that all multipartite entanglement witnesses can be constructed from the solutions of the MSEvalue equations,
	\begin{align}
		\hat W=\sup\{g\}\hat 1-\hat L.
	\end{align}
	In case we consider finite Hilbert spaces ($\dim\mathcal H<\infty$), we can replace $\sup$ and $\inf$ by $\max$ and $\min$, respectively.


	The derived MSEvalue equations play a fundamental role for multipartite entanglement tests.
	They give the possibility to construct arbitrary entanglement witnesses on the basis of the solution of an algebraic eigenvalue problem of an observable $\hat L$.
	Numerical and analytical methods -- originally developed to solve eigenvalue problems -- can be applied to handle the multipartite entanglement problem in quantum physics in a systematic way.
	Before we apply our method, let us formulate some fundamental properties of the MSEvalue equations.
	The proofs are given in the Appendix.

	\paragraph{Proposition: Second form of MSEvalue equations.--}
	The Hermitian operator $\hat L$ has the MSEvalue $g$ for the MSEvector $|a_1,\dots,a_K\rangle$, if and only if it fulfills the second form of the MSEvalue equations
	\begin{align*}
		\hat L|a_1,\dots,a_K\rangle=g|a_1,\dots,a_K\rangle+|\chi\rangle,
	\end{align*}
	with $\langle a_1,\dots,a_{j-1},x,a_{j+1},\dots,a_K|\chi\rangle=0$ for all $|x\rangle\in\bigotimes_{i\in\mathcal I_j}\mathcal H_i$ and $j=1,\dots,K$.
	\hfill$\blacksquare$

	This proposition transforms the coupled system of eigenvalue equations, which has been defined in the first form of the MSEvalue equations, into a single, but perturbed eigenvalue problem.
	This second form yields several implications.
	For example, if an eigenvector is a product vector $|a_1,\dots,a_K\rangle$, it is also an MSEvector with $|\chi\rangle\equiv0$.
	In addition, we also conclude that the operator $\hat L$ yields a true entanglement witness, cf. Eq.~\eqref{Eq:TrueWitness}, if and only if the eigenspace of the largest eigenvalue, does not contain a product vector.

	\paragraph{Proposition: Transformation properties.--}
	A Hermitian operator $\hat L$ has a MSEvalue $g$ for the MSEvector $|a_1,\dots,a_K\rangle$.
	Then, the operator
	\begin{align*}
		\hat L'= \left(\hat U_1\otimes\dots\otimes \hat U_K\right)^\dagger
		\left[\lambda_1\hat 1+\lambda_2 \hat L\right]
		\left(\hat U_1\otimes\dots\otimes \hat U_K\right),
	\end{align*}
	with $\lambda_1,\lambda_2\in\mathbb R\setminus \{0\}$ and $\hat U_j$ being unitary operations acting locally on the partition $\mathcal I_j$,
	has the MSEvalue $g'=\lambda_1+\lambda_2 g$ and the MSEvectors
	$|a'_1,\dots,a'_K\rangle=\left.\hat U_1^\dagger\otimes\dots\otimes \hat U_K^\dagger\right.|a_1,\dots,a_K\rangle$.
	\hfill$\blacksquare$

	This transformation allows us to consider a whole class of witnesses, by solving the MSEvalue equation for a particular operator $\hat L$.
	In addition, the shifting of $\hat L$ to $\hat L'$ allows us to consider positive semidefinite operators only.
	Note that the invariance of the MSEvalues under local unitaries is of particular interest for the quantification of multipartite entanglement, see, e.g.,~\cite{horodecki,guehne}.

	\paragraph{Proposition: Cascaded structure.--}
	The nonzero solutions of an $N+1$-partite operator $\hat L'=|\psi\rangle\langle\psi|$ are identical to the solutions of an $N$-partite operator $\hat L={\rm tr}_{N+1} \hat L'$.
	\hfill$\blacksquare$

	This property is quite surprising.
	It shows us that all possible entanglement witnesses -- based on positive semidefinite operators $\hat L$ -- of an $N$-partite system can be constructed by a few simple entanglement witnesses in a $N+1$-partite system, $\hat L'=|\psi\rangle\langle\psi|$.
	An arbitrary rank of $\hat L$ can be achieved by choosing a state $|\psi\rangle$ with the same Schmidt rank for the bipartition $\mathcal I_1=\{1,\dots,N\}$ and $\mathcal I_2=\{N+1\}$, cf. the Appendix.


	In the following, we apply our method to analytically derive multipartite entanglement tests.
	First, we may consider witnesses for prominent examples of states in a three qubit systems.
	In a second step, we apply our method to get a multipartite entanglement test in a complex continuous variable system.

\begin{figure}
	\includegraphics*[height=4cm]{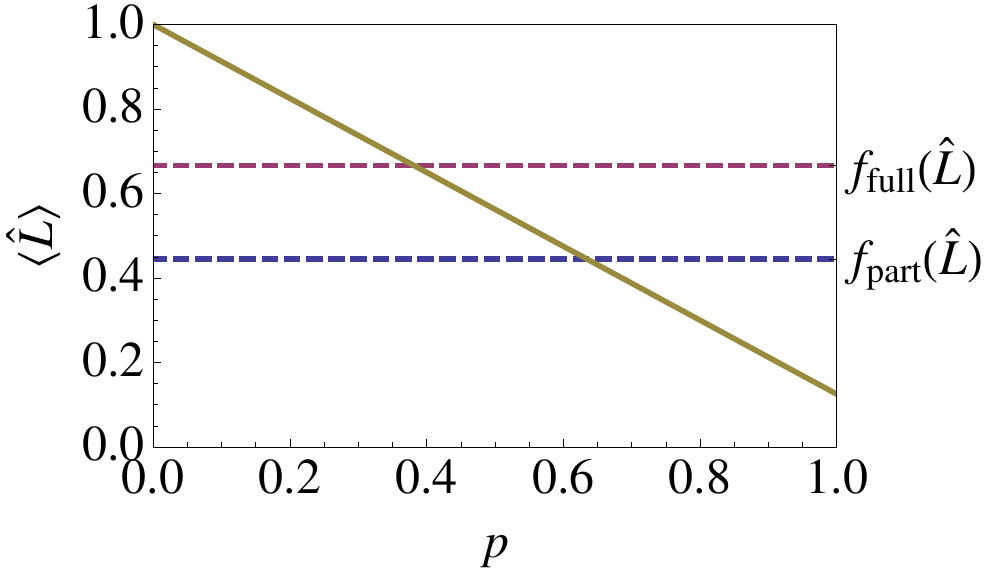}
	\caption{
		(Color online)
		The entanglement test in Eq.~\eqref{Eq:Supremum} for a W-state mixed with white noise, $\hat \rho=p\frac{1}{8}\hat 1+(1-p)|\psi_{\rm W}\rangle\langle\psi_{\rm W}|$ $\lambda_1=\lambda_2=\lambda_3=\frac{1}{\sqrt 3}$, is plotted for $0\leq p\leq 1$.
		The boundary for partial or full separability is $\frac{4}{9}$ or $\frac{2}{3}$, respectively.
		The expectation value $\langle \hat L\rangle$ exceeds the boundary for full or partial separability as long as the mixing parameter is $p<\frac{40}{63}$ or $p<\frac{8}{21}$, respectively.
	}\label{Fig:Wstate}
	\end{figure}

	Let us consider a generalized tripartite W-state
	\begin{align}
		|\psi_{\rm W}\rangle=\lambda_1|1,0,0\rangle+\lambda_2|0,1,0\rangle+\lambda_3|0,0,1\rangle,
	\end{align}
	with $|\lambda_1|^2+|\lambda_2|^2+|\lambda_3|^2=1$, which defines the observable $\hat L=|\psi_{\rm W}\rangle\langle\psi_{\rm W}|$.
	In the Appendix, we solve the MSEvalue equation of $\hat L$. This gives
	\begin{align}
		\nonumber f_{\{1\}:\{2,3\}}(\hat L)=&\max\{|\lambda_1|^2,|\lambda_2|^2+|\lambda_3|^2\}, \\
		\nonumber f_{\{2\}:\{1,3\}}(\hat L)=&\max\{|\lambda_2|^2,|\lambda_1|^2+|\lambda_3|^2\}, \\
		\nonumber f_{\{3\}:\{1,2\}}(\hat L)=&\max\{|\lambda_3|^2,|\lambda_1|^2+|\lambda_2|^2\}, \\
		f_{\{1\}:\{2\}:\{3\}}(\hat L)=&\max\{|\lambda_1|^2,|\lambda_2|^2,|\lambda_3|^2,g_{0}\},
	\end{align}
	with
	\begin{align}
		\nonumber g_{0}= \frac{4|\lambda_1|^2|\lambda_2|^2|\lambda_3|^2}{(|\lambda_1|^2+|\lambda_2|^2+|\lambda_3|^2)^2-2(|\lambda_1|^4+|\lambda_2|^4+|\lambda_3|^4)}.
	\end{align}
	Hence, we can formulate the following multipartite entanglement conditions: 
	A quantum state $\hat \rho$ is partially entangled, if $\langle\psi_{\rm W}|\hat \rho|\psi_{\rm W}\rangle>f_{\{1\}:\{2\}:\{3\}}(\hat L)$.
	The corresponding entanglement witness is
	\begin{align}
		\hat W_{\rm part}=\max\{g_{0},|\lambda_1|^2,|\lambda_2|^2,|\lambda_3|^2\}\hat 1-|\psi_{\rm W}\rangle\langle\psi_{\rm W}|.
	\end{align}
	A quantum state $\hat \rho$ is fully entangled, if $\langle\psi_{\rm W}|\hat\rho|\psi_{\rm W}\rangle>\max\{f_{\{1\}:\{2,3\}}(\hat L),f_{\{2\}:\{1,3\}}(\hat L),f_{\{3\}:\{1,2\}}(\hat L)\}$.
	The corresponding entanglement witness is
	\begin{align}
		\hat W_{\rm full}=\max\{|\lambda_i|^2+|\lambda_j|^2: i\neq j\}\hat 1-|\psi_{\rm W}\rangle\langle\psi_{\rm W}|.
	\end{align}
	In Fig.~\ref{Fig:Wstate}, we apply the considered witness to study the entanglement of a noisy W-state.

\begin{figure}
	\includegraphics*[height=4cm]{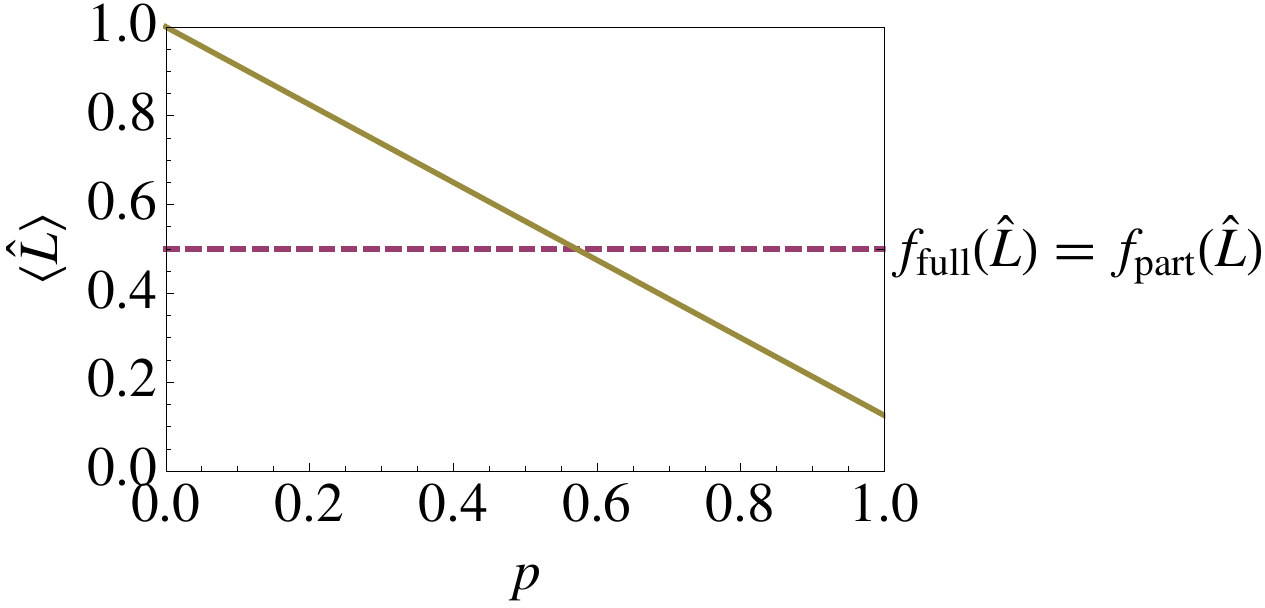}
	\caption{
		(Color online)
		The expectation value $\langle\hat L\rangle$ for a GHZ-state mixed with white noise, $\hat\rho=p\frac{1}{8}\hat 1+(1-p)|\psi_{\rm GHZ}\rangle\langle\psi_{\rm GHZ}|$ ($\kappa_0=\kappa_1=\frac{1}{\sqrt 2}$), is plotted.
		The lower boundary for partial and full entanglement is $\frac{1}{2}$.
		The expectation value $\langle \hat L\rangle$ exceeds this boundary -- which automatically implies genuine entanglement -- as long as the mixing parameter is $p<\frac{8}{14}$.
	}\label{Fig:GHZstate}
	\end{figure}

	In a second step, a generalized GHZ-state is given,
	\begin{align}
		|\psi_{\rm GHZ}\rangle=\kappa_0|0,0,0\rangle+\kappa_1|1,1,1\rangle,
	\end{align}
	together with $|\kappa_0|^2+|\kappa_1|^2=1$, which yields an observable $\hat L=|\psi_{\rm GHZ}\rangle\langle\psi_{\rm GHZ}|$.
	From the Appendix, we get the maximal MSEvalues:
	\begin{align}
		\nonumber &f_{\{1\}:\{2,3\}}(\hat L)=f_{\{2\}:\{1,3\}}(\hat L)=f_{\{3\}:\{1,2\}}(\hat L)\\
		=&f_{\{1\}:\{2\}:\{3\}}(\hat L)=\max\{|\kappa_0|^2,|\kappa_1|^2\}.\label{Eq:fGHZ}
	\end{align}
	Hence, a state $\hat \rho$ is genuinely tripartite entangled, if $\langle\psi_{\rm GHZ}|\hat \rho|\psi_{\rm GHZ}\rangle>\max\{|\kappa_0|^2,|\kappa_1|^2\}$, see Fig.~\ref{Fig:GHZstate}.
	Note that the corresponding witness 
 	\begin{align}       
		\hat W=\max\{|\kappa_0|^2,|\kappa_1|^2\}\hat 1-|\psi_{\rm GHZ}\rangle\langle\psi_{\rm GHZ}|
        \end{align}
        cannot discriminate between partially and fully entangled states, see Eq.~\eqref{Eq:fGHZ}, which is possible for the generalized W-state projection in the previous example.


	As a proof of principle, we are going to test multipartite entanglement of a continuous variable system.
	Our considered example is a system of $N$ coupled harmonic oscillators.
	The observable we are using to verify entanglement is the total energy of this system, $\hat L=\hat H$,
	\begin{align}
		\hat H=\sum_{j=1}^N \left( \frac{{\vec{\hat p}_j}^2}{2m}+\frac{m\omega^2{\vec{\hat r}_j}^2}{2}\right)+\frac{\gamma}{4}\sum_{j,j'=1}^N|\vec{\hat r}_j-\vec{\hat r}_{j'}|^2,
	\end{align}
	where $\gamma$ denotes the coupling strength of the interaction, $\vec{\hat r}_j$ the position and $\vec{\hat p}_j$ the momentum operator.
	Let us note that we considered an even more general case in the Appendix.
	For the partition $\mathcal I_1,\dots,\mathcal I_K$, we get the smallest MSEvalue of the Hamiltonian as
	\begin{align}
		\nonumber E[\mathcal I_1,\dots,\mathcal I_K]=\frac{3}{2}\hbar\omega&\sum_{j=1}^K\left([N_j-1]\sqrt{1+N\frac{\gamma}{m\omega^2}}\right.\\
		&+\left.\sqrt{1+[N-N_j]\frac{\gamma}{m\omega^2}}\,\right),
	\end{align}
	where $N_j=|\mathcal I_j|$ is the number of subsystems in $\mathcal I_j$.
	The resulting witness reads as
	\begin{align}       
		\hat W_{\mathcal I_1,\dots,\mathcal I_K}= E[\mathcal I_1,\dots,\mathcal I_K]\hat 1- \hat H.
        \end{align}
 
\begin{figure}
	\includegraphics*[height=5.5cm]{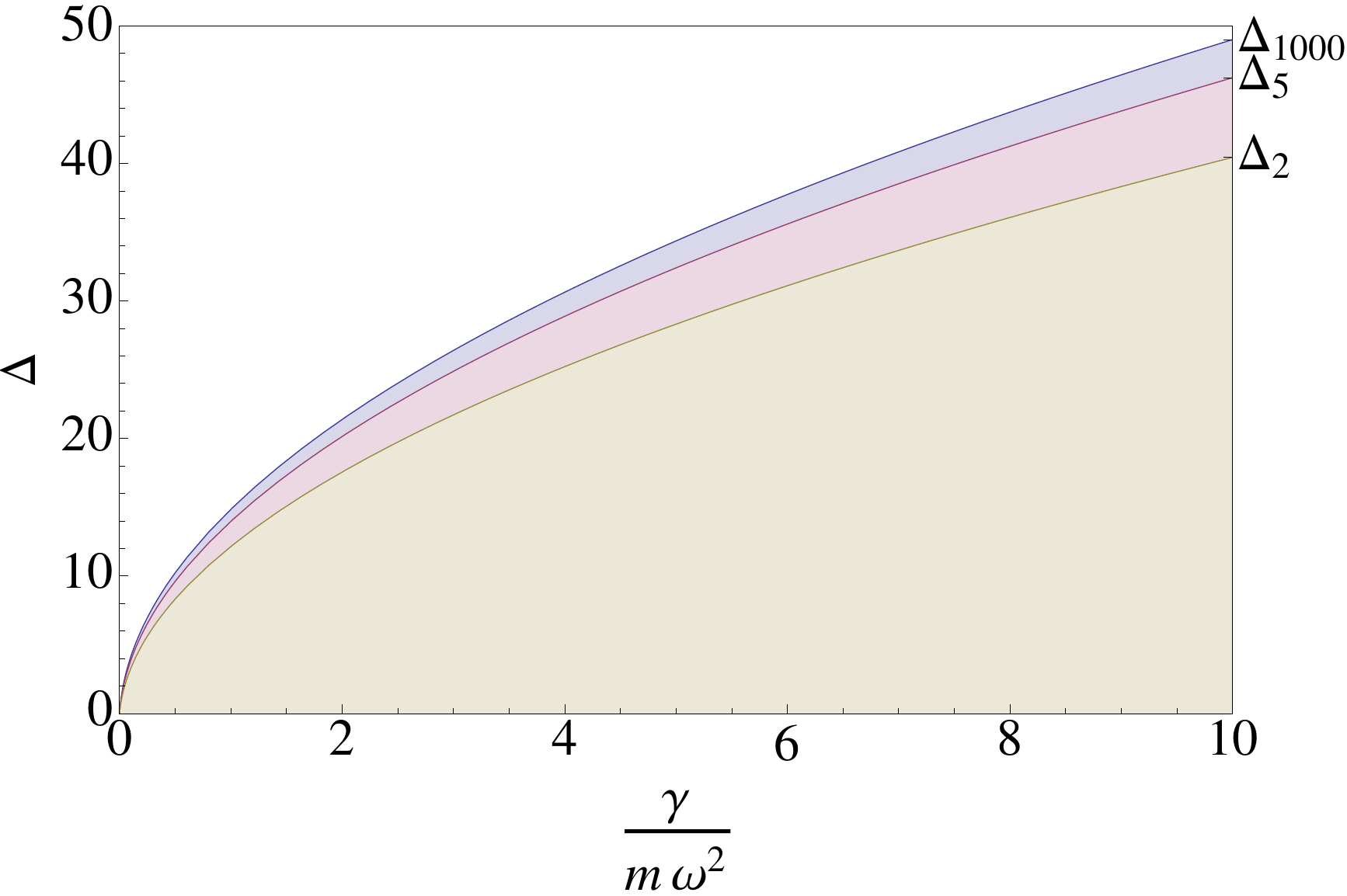}
	\caption{
		(Color online)
		The energy difference $\Delta=(\langle\hat H\rangle-E[\mathcal I])/\left(\frac{3}{2}\hbar\omega\right)$ between quantum states and the ground state of the system is plotted depending on the interaction strength.
		The boundary $\Delta_{1000}$ corresponds to the minimal attainable energy for fully separable states.
		All quantum states having an energy below $\Delta_5$ ($N_1=\dots=N_5=200$) reveal entanglement between more than 200 oscillators.
		The quantum states in the range below $\Delta_2$ ($N_1=N_2=500$) exhibit entanglement distributed over more than 500 subsystems.
	}\label{Fig:Gaussianstate}
	\end{figure}

	In the special case $K=1$ ($\mathcal I_1=\mathcal I$) we get the true ground state energy of the system,
	\begin{align}
		E[\mathcal I]=\frac{3}{2}\hbar\omega\left(\sqrt{1+N\frac{\gamma}{m\omega^2}}(N-1)+1\right).
	\end{align}
	In case of full separability, $K=N$ ($\mathcal I_j=\{j\}$), we have a minimal energy of:
	\begin{align}
		E[\{1\},\dots,\{N\}]=\frac{3}{2}\hbar\omega N\sqrt{1+[N-1]\frac{\gamma}{m\omega^2}}.
	\end{align}
	In Fig.~\ref{Fig:Gaussianstate}, we plotted the corresponding entanglement test based on Eq.~\eqref{Eq:Infimum}, for $N=10^3$ interacting oscillators.
	For the witnessing by the total energy $\hat H$, no information about the structure of the quantum states is needed.

	In conclusion, we have derived an algebraic set of equations to construct arbitrary entanglement witnesses.
	We studied some fundamental properties of these equations.
	For example, they are invariant under local unitary transformations and have a cascaded structure.
	The latter allows us to deduce all entanglement witnesses from elementary projections.
	Our method enables us to use all known procedures for solving eigenvalue problems to construct entanglement witnesses.
	We applied our method to analytically identify full and partial entanglement of generalized, noisy GHZ- and W-states.
	Moreover, we witnessed multipartite entanglement for a system of $10^3$ interacting oscillators, by analytical computation of the energetic boundaries of separable states.
	This demonstrates the feasibility of our method for studying ultrahigh orders of multipartite entanglement, which are of fundamental interest for understanding the transition from the microscopic to the macroscopic world.

	The authors are grateful to D. Pagel for helpful comments.
	This work was supported by the Deutsche Forschungsgemeinschaft through SFB 652.

\appendix
\begin{widetext}

\section{Supplemental Material -- Multipartite Entanglement Witnesses}
	Includes the proofs of the propositions, and the analytical calculation of the MSEvalues for the considered examples.

\subsection{Proof: Second Form of MSEvalue equations}
	Let us assume that $g$ and $|a_1,\dots,a_K\rangle$ are a MSEvalue and a MSEvector of $\hat L$, respectively.
	This means they fulfill $\hat L_{a_1,\dots,a_{j-1},a_{j+1},a_K}|a_j\rangle=g|a_j\rangle$ for all $j=1,\dots,K$.
	The result of the mapping of $\hat L$ acting on a vector $|a_1,\dots,a_K\rangle$ can be decomposed in one part parallel to the input state and another part $|\chi\rangle$ perpendicular to the input state
	\begin{align}
		\hat L|a_1,\dots,a_K\rangle=&g'|a_1,\dots,a_K\rangle+|\chi\rangle
	\end{align}
	Note that $g=\langle a_1,\dots,a_K|\hat L|a_1,\dots,a_K\rangle=g'$.
	Let us consider an arbitrary $|x\rangle\in\bigotimes_{i\in\mathcal I_j}\mathcal H_i$ ($j=1,\dots,K$).
	We get the following projection for the map $\hat L-g\hat 1$:
	\begin{align}
		\nonumber 0=&g-g=\langle x|(\hat L-g\hat 1)_{a_1,\dots,a_{j-1},a_{j+1},a_K}|a_j\rangle\\
		\nonumber =&\langle a_1,\dots,a_{j-1},x,a_{j+1},a_K|\underbrace{\hat L|a_1,\dots,a_K\rangle}_{=g|a_1,\dots,a_K\rangle+|\chi\rangle}-g\langle a_1,\dots,a_{j-1},x,a_{j+1},a_K|a_1,\dots,a_K\rangle\\
		\nonumber =&g\langle a_1,\dots,a_{j-1},x,a_{j+1},a_K|a_1,\dots,a_K\rangle+\langle a_1,\dots,a_{j-1},x,a_{j+1},a_K|\chi\rangle\\
		\nonumber &-g\langle a_1,\dots,a_{j-1},x,a_{j+1},a_K|a_1,\dots,a_K\rangle\\
		=&\langle a_1,\dots,a_{j-1},x,a_{j+1},a_K|\chi\rangle.\label{Eq:Northogonal}
	\end{align}
	Hence, the MSEvalue equations imply that for the solution holds the form
	\begin{align}\label{Eq:SecondForm}
		\hat L|a_1,\dots,a_K\rangle=&g|a_1,\dots,a_K\rangle+|\chi\rangle,
	\end{align}
	where $|\chi\rangle$ has the property given in Eq.~\eqref{Eq:Northogonal}.
	The other way around, we get from these equivalent transformations, that Eq.~\eqref{Eq:SecondForm} implies that $g$ and $|a_1,\dots,a_K\rangle$ are a MSEvalue and a MSEvector of $\hat L$, respectively.

\subsection{Proof: Transformation Properties}
	Let us assume that $g$ and $|a_1,\dots,a_K\rangle$ are a MSEvalue and a MSEvector of $\hat L$, respectively.
	Hence, Eq.~\eqref{Eq:SecondForm} is fulfilled.
	Now, we consider an operator $\hat L'$ ($\lambda_1,\lambda_2\in\mathbb R\setminus \{0\}$):
	\begin{align}
		\hat L'=\left(\hat U_1\otimes\dots\otimes \hat U_K\right)^\dagger\left[\lambda_1\hat 1+\lambda_2 \hat L\right]\left(\hat U_1\otimes\dots\otimes \hat U_K\right),
	\end{align}
	where $\hat U_j$ are unitary transformations acting on $\bigotimes_{i\in\mathcal I_j}\mathcal H_i$ ($j=1,\dots,K$).
	We get for the transformed input vector
	\begin{align}
		\nonumber \hat L'|a'_1,\dots,a'_K\rangle=&\hat L'\left(\hat U_1^\dagger |a_1\rangle \otimes\dots\otimes \hat U_K^\dagger |a_K\rangle\right)\\
		\nonumber =&\left(\hat U_1\otimes\dots\otimes \hat U_K\right)^\dagger\left(\lambda_1|a_1,\dots,a_K\rangle+\lambda_2 g |a_1,\dots,a_K\rangle +|\chi\rangle\right)\\
		=&(\lambda_1+\lambda_2 g)|a'_1,\dots,a'_K\rangle+\left(\hat U_1\otimes\dots\otimes \hat U_K\right)^\dagger|\chi\rangle
		=(\lambda_1+\lambda_2 g)|a'_1,\dots,a'_K\rangle+|\chi'\rangle,\label{Eq:TransformedSE}
	\end{align}
	where $|\chi'\rangle$ obviously fulfills Eq.~\eqref{Eq:Northogonal} for $|a'_1,\dots,a'_K\rangle$ and arbitrary $|x'\rangle=\hat U_j^\dagger|x\rangle\in\bigotimes_{i\in\mathcal I_j}\mathcal H_i$.
	Hence, Eq.~\eqref{Eq:TransformedSE} is the second form of the MSEvalue equation for $\hat L'$ with the transformed solutions:
	\begin{align}
		g'=\lambda_1+\lambda_2 g \text{ and } |a'_1,\dots,a'_K\rangle=\hat U_1^\dagger |a_1\rangle \otimes\dots\otimes \hat U_K^\dagger |a_K\rangle.
	\end{align}
	Let us note that an inverse transformation of $\hat L'$ yields the inverse implication,
	\begin{align}
		\hat L=\left(\hat U_1\otimes\dots\otimes \hat U_K\right)\left[-\frac{\lambda_1}{\lambda_2}\hat 1+\frac{1}{\lambda_2} \hat L'\right]\left(\hat U_1\otimes\dots\otimes \hat U_K\right)^\dagger.
	\end{align}

\subsection{Proof: Cascaded Structure}
	We consider a rank one operator $\hat L'=|\psi\rangle\langle\psi|$.
	Let us consider the Schmidt decomposition~\cite{nielsenchuang} of $|\psi\rangle$, with respect to a bipartite decomposition $\{1,\dots,N\}$ and  $\{N+1\}$, as
	\begin{align}
		|\psi\rangle=\sum_{j=1}^r \lambda_j |\psi_j\rangle\otimes |z_j\rangle,
	\end{align}
	with $\lambda_j\in\mathbb R_{>0}$ (Schmidt coefficients), $r$ (Schmidt rank), orthonormal $|\psi_j\rangle\in\bigotimes_{i=1}^{N}\mathcal H_i$, and orthonormal $|z_j\rangle\in\mathcal H_{N+1}$.
	The MSEvalue equation in the first form for the $N+1$-th subsystem reads as
	\begin{align}\label{Eq:SENthParty}
		g|a_{N+1}\rangle=&\hat L'_{a_1,\dots,a_{N}}|a_{N+1}\rangle=\langle\psi|a_1,\dots,a_{N+1}\rangle\sum_{j=1}^r\lambda_j\langle a_1,\dots,a_{N}|\psi_j\rangle |z_j\rangle,
	\end{align}
	with the abbreviation for the MSEvalue $g=|\gamma|^2$ ($\gamma=\langle\psi|a_1,\dots,a_{N+1}\rangle$).
	Here, we are only interested in solutions with $g\neq0$.
	From Eq.~\eqref{Eq:SENthParty}, we get that the solution $|a_{N+1}\rangle$ must have the form
	\begin{align}
		|a_{N+1}\rangle=\sum_{j=1}^r\frac{\lambda_j\langle a_1,\dots,a_{N}|\psi_j\rangle}{\gamma^\ast}|z_j\rangle.
	\end{align}
	We may insert this result into the MSEvalue equation for the $i$-th party ($i=1,\dots,N$):
	\begin{align}
		\nonumber g|a_i\rangle=&\hat L'_{a_1,\dots,a_{i-1},a_{i+1},\dots a_{N+1}}|a_i\rangle\\
		\nonumber =&\langle\psi|a_1,\dots,a_{N+1}\rangle\sum_{j=1}^r\lambda_j\langle a_{N+1}|z_j\rangle\left(\langle a_1,\dots,a_{i-1}|\otimes\,\cdot\,\otimes\langle a_{i+1},\dots,a_{N}|\right) |\psi_j\rangle\\
		\nonumber =&\gamma\sum_{j=1}^r \frac{\lambda_j^2}{\gamma}\left(\langle a_1,\dots,a_{i-1}|\otimes\,\cdot\,\otimes\langle a_{i+1},\dots,a_{N}|\right) |\psi_j\rangle \langle\psi_j|a_1,\dots,a_{N}\rangle\\
		\nonumber =&\left(\langle a_1,\dots,a_{i-1}|\otimes\,\cdot\,\otimes\langle a_{i+1},\dots,a_{N}|\right)
		\left(\sum_{j=1}^r \lambda_j^2|\psi_j\rangle\langle\psi_j|\right)
		\left(|a_1,\dots,a_{i-1}\rangle\otimes\,|a_i\rangle\otimes| a_{i+1},\dots,a_{N}\rangle\right)\\
		=& \hat L_{a_1,\dots,a_{i-1},a_{i+1},\dots,a_{N}} |a_i\rangle,
	\end{align}
	where we used the abbreviation
	\begin{align}
		\hat L={\rm tr}_{N+1} |\psi\rangle\langle\psi|=\sum_{j,j'=1}^r \lambda_j\lambda_{j'}{\rm tr}_{N+1}(|z_j\rangle\langle z_{j'}|)|\psi_j\rangle\langle\psi_{j'}|
		=\sum_{j=1}^r \lambda_j^2 |\psi_j\rangle\langle\psi_j|.
	\end{align}
	Let us note that $\hat L$ is a positive semi-definite operator which has a rank $r$.

\subsection{W-state witness}
	Now, we consider the operator $\hat L=|\psi_{\rm W}\rangle\langle\psi_{\rm W}|$, where $|\psi_{\rm W}\rangle=\lambda_1|1,0,0\rangle+\lambda_2|0,1,0\rangle+\lambda_3|0,0,1\rangle$ is a generalized W-state.
	The reduced operator -- with respect to the third subsystem -- is
	\begin{align}
		\hat L'={\rm tr}_3\, \hat L=|\psi'\rangle\langle\psi'|+|\lambda_3|^2|0,0\rangle\langle 0,0|,\text{ with }|\psi'\rangle=\lambda_1|1,0\rangle+\lambda_2|0,1\rangle.
	\end{align}
	We may start with the calculation of the maximal MSEvalue for partial separability in the case $\{3\}:\{1,2\}$.
	Since $|\psi\rangle'$ and $|0,0\rangle$ are perpendicular vectors in $\mathcal H_1\otimes\mathcal H_2$, we immediately get
	\begin{align}
		g_{\max}=\max\{|\lambda_3|^2,\langle\psi'|\psi'\rangle=|\lambda_1|^2+|\lambda_2|^2\}.
	\end{align}
	Similarly, this yields the maximal MSEvalues for the other decompositions: $g_{\max}=\max\{|\lambda_1|^2,|\lambda_2|^2+|\lambda_3|^2\}$ for $\{1\}:\{2,3\}$ and $g_{\max}=\max\{|\lambda_2|^2,|\lambda_1|^2+|\lambda_3|^2\}$ for $\{2\}:\{1,3\}$.

	For the full separable case $\{1\}:\{2\}:\{3\}$, we can easily get the obvious solutions of $\hat L'$:
	\begin{align}\label{Eq:WstateObvious}
		g=|\lambda_3|^2\text{ for }|a_1,a_2\rangle=|0,0\rangle,\,
		g=|\lambda_1|^2\text{ for }|a_1,a_2\rangle=|1,0\rangle,\\
		g=|\lambda_2|^2\text{ for }|a_1,a_2\rangle=|0,1\rangle,
		\text{ and }g=0\text{ for }|a_1,a_2\rangle=|1,1\rangle.\nonumber
	\end{align}
	A convenient parametrization of all other possible MSEvectors is
	\begin{align}\label{Eq:parametrization}
		|a_1,a_2\rangle=\frac{|0\rangle+\alpha_1|1\rangle}{\sqrt{1+|\alpha_1|^2}}\otimes\frac{|0\rangle+\alpha_2|1\rangle}{\sqrt{1+|\alpha_2|^2}},\text{ for $\alpha_1,\alpha_2\in\mathbb C\setminus\{0\}$.}
	\end{align}
	Here, we use the MSEvalue equations in the second form $\hat L'|a_1,a_2\rangle=g|a_1,a_2\rangle+|\chi\rangle$.
	Since we require that $\langle x_1,a_2|\chi\rangle=0$ and $\langle a_1,x_2|\chi\rangle=0$ for all $|x_i\rangle\in\mathcal H_i$ ($i=1,2$), we have
	\begin{align}
		|\chi\rangle=h|a_1^\perp,a_2^\perp\rangle=h\frac{-\alpha_1^\ast|0\rangle+|1\rangle}{\sqrt{1+|\alpha_1|^2}}\otimes\frac{-\alpha_2^\ast|0\rangle+|1\rangle}{\sqrt{1+|\alpha_2|^2}}
		\text{ for some $h\in\mathbb C$}.
	\end{align}
	This leads to the MSEvalue equation in the second form as
	\begin{align}
		\nonumber &|\lambda_3||0,0\rangle
		+\lambda_2(\lambda_2^\ast\alpha_2+\lambda_1^\ast\alpha_1)|0,1\rangle
		+\lambda_1(\lambda_2^\ast\alpha_2+\lambda_1^\ast\alpha_1)|1,0\rangle
		+0|1,1\rangle\\
		=&(g+h\alpha_1^\ast\alpha_2^\ast)|0,0\rangle
		+(g\alpha_2-h\alpha_1^\ast)|0,1\rangle
		+(g\alpha_1-h\alpha_2^\ast)|1,0\rangle
		+(g\alpha_1\alpha_2+h)|1,1\rangle,
	\end{align}
	where we multiplied the whole equation with the normalization constant $\sqrt{1+|\alpha_1|^2}\sqrt{1+|\alpha_2|^2}$.
	We may decompose $\alpha_i=r_i{\rm e}^{{\rm i}\varphi_i}$ for $i=1,2$ in polar coordinates.
	The individual basis components can be rewritten as
	\begin{align}
		\label{Eq:Component00} |\lambda_3|^2=&g+h'r_1r_2\\
		\label{Eq:Component01} \lambda_2'(\lambda_2'^\ast r_2+\lambda_1'^\ast r_1)=&gr_2-h'r_1\\
		\label{Eq:Component10} \lambda_1'(\lambda_2'^\ast r_2+\lambda_1'^\ast r_1)=&gr_1-h'r_2\\
		\label{Eq:Component11} 0=&g r_1r_2 +h',
	\end{align}
	where we introduced $h'=h{\rm e}^{-{\rm i}(\varphi_1+\varphi_2)}$ and $\lambda_i'=\lambda_i{\rm e}^{-{\rm i}\varphi_i}$ ($i=1,2$).
	Since $g=\langle a_1,a_2|\hat L'|a_1,a_2\rangle\in\mathbb R$, it follows from Eq.~\eqref{Eq:Component11}, that $h'\in\mathbb R$.
	Eqs.~\eqref{Eq:Component01}~and~\eqref{Eq:Component10} can be combined to
	\begin{align}
		\begin{pmatrix}
			|\lambda_2'|^2-g & \lambda_2'\lambda_1'^\ast+h' \\
			\lambda_1'\lambda_2'^\ast+h' & |\lambda_1'|^2-g \\
		\end{pmatrix}
		\begin{pmatrix}r_2 \\ r_1\end{pmatrix}
		=\begin{pmatrix}0\\0\end{pmatrix}.
	\end{align}
	Note that this equation has a real solution $(r_1,r_2)$, iff ${\rm Im}(\lambda_2'\lambda_1'^\ast)=0$ or, equivalently, $\lambda_2'\lambda_1'^\ast=\lambda_1'\lambda_2'^\ast$.
	More generally, this equation has a solution, iff
	\begin{align}\label{Eq:Quadratic1}
		0=\det\begin{pmatrix}
			|\lambda_2'|^2-g & \lambda_2'\lambda_1'^\ast+h' \\
			\lambda_1'\lambda_2'^\ast+h' & |\lambda_1'|^2-g \\
		\end{pmatrix}
		=(|\lambda_1|^2-g)(|\lambda_2|^2-g)-|\lambda_2'\lambda_1'^\ast+h'|^2.
	\end{align}
	In addition, we may resolve Eqs.~\eqref{Eq:Component00}~and~\eqref{Eq:Component11} as
	\begin{align}\label{Eq:Quadratic2}
		\frac{|\lambda_3|^2-g}{h'}=r_1r_2=\frac{-h'}{g}
		\Leftrightarrow
		h'^2=g(g-|\lambda_3|^2).
	\end{align}
	Hence, we get quadratic equations~\eqref{Eq:Quadratic1}~and~\eqref{Eq:Quadratic2} in $g$ and $h'$.
	The difference of both equations gives the relation between $g$ and $h'$ as
	\begin{align}
		h'=\frac{|\lambda_3|^2-|\lambda_1|^2-|\lambda_2|^2}{2\lambda_2'\lambda_1'^\ast}g.
	\end{align}
	This relation, we may insert into Eq.~\eqref{Eq:Quadratic2} which results in
	\begin{align}
		0=g\left(\left[1-\frac{(|\lambda_3|^2-|\lambda_1|^2-|\lambda_2|^2)^2}{4|\lambda_1|^2|\lambda_2|^2}\right]g-|\lambda_3|^2\right).
	\end{align}
	Omitting $g=0$, we get the root of this equation as
	\begin{align}
		\nonumber g_0=&\frac{4|\lambda_1|^2|\lambda_2|^2|\lambda_3|^2}{2(|\lambda_1|^2|\lambda_2|^2+|\lambda_1|^2|\lambda_3|^2+|\lambda_2|^2|\lambda_3|^2)-|\lambda_1|^4-|\lambda_2|^4-|\lambda_3|^4}\\
		=&\frac{4|\lambda_1|^2|\lambda_2|^2|\lambda_3|^2}{(|\lambda_1|^2+|\lambda_2|^2+|\lambda_3|^2)^2-2(|\lambda_1|^4+|\lambda_2|^4+|\lambda_3|^4)}.
	\end{align}
	Finally, we conclude for the case of full separability that the maximal MSEvalue $g_{\max}$ of $\hat L$ is
	\begin{align}
		g_{\max}=\max\{|\lambda_1|^2,|\lambda_2|^2,|\lambda_3|^2,g_{0}\},
	\end{align}
	where we have taken the solutions in Eq.~\eqref{Eq:WstateObvious} into account.
	It is of importance to mention that from the calculation of the MSEvector of $g_0$ follows, that we have to fulfill the requirements $|\lambda_1|^2+|\lambda_2|^2\geq|\lambda_3|^2$, $|\lambda_3|^2+|\lambda_2|^2\geq|\lambda_1|^2$, and $|\lambda_1|^2+|\lambda_3|^2\geq|\lambda_2|^2$.

\subsection{GHZ-state witness}
	Let us consider the operator $\hat L=|\psi_{\rm GHZ}\rangle\langle\psi_{\rm GHZ}|$, with $|\psi_{\rm GHZ}\rangle=\kappa_0|0,0,0\rangle+\kappa_1|1,1,1\rangle$ being a generalized GHZ-state.
	The reduced operator is
	\begin{align}\label{Eq:30}
		\hat L'={\rm tr}_3\, \hat L=|\kappa_0|^2 |0,0\rangle\langle0,0|+|\kappa_1|^2 |1,1\rangle\langle1,1|.
	\end{align}
	First, we may consider a decomposition $\{3\}:\{1,2\}$.
	Since $\hat L'$ is already given in a spectral decomposition, we get the non-zero MSEvalues $|\kappa_0|^2$ and $|\kappa_1|^2$.
	Similarly, we can argue for the partial separability with respect to $\{2\}:\{1,3\}$ and $\{1\}:\{2,3\}$.
	We get that the maximal MSEvalue, $g_{\max}$, for partial separability is
	\begin{align}
		g_{\max}=\max\{|\kappa_0|^2,|\kappa_1|^2\}.
	\end{align}

	Second, we consider full separability $\{1\}:\{2\}:\{3\}$.
	Due to the reduction from $\hat L$ to $\hat L'$, we have to solve the MSEvalue equations of $\hat L'$ as given in Eq.~\eqref{Eq:30}.
	It is easy to check that $|i,i\rangle$ is an MSEvector for the MSEvalue $|\kappa_i|^2$ ($i=0,1$).
	For all other MSEvectors, we use the same parametrization as given in Eq.~\eqref{Eq:parametrization}.
	Now, the MSEvalue equation read as
	\begin{align}
		\hat L'_{a_1}|a_2\rangle=g|a_2\rangle &\Leftrightarrow
		\frac{|\kappa_0|^2|0\rangle+|\kappa_1|^2|\alpha_1|^2\alpha_2|1\rangle}{(1+|\alpha_1|^2)\sqrt{1+|\alpha_2|^2}}=g\frac{|0\rangle+\alpha_2|1\rangle}{\sqrt{1+|\alpha_2|^2}}\\
		\hat L'_{a_2}|a_1\rangle=g|a_1\rangle &\Leftrightarrow
		\frac{|\kappa_0|^2|0\rangle+|\kappa_1|^2|\alpha_2|^2\alpha_1|1\rangle}{(1+|\alpha_2|^2)\sqrt{1+|\alpha_1|^2}}=g\frac{|0\rangle+\alpha_1|1\rangle}{\sqrt{1+|\alpha_1|^2}}
	\end{align}
	These equations are fulfilled, if $|\alpha_1|=|\alpha_2|=|\kappa_0|/|\kappa_1|$ for the MSEvalue $g=|\kappa_0|^2|\kappa_1|^2/(|\kappa_0|^2+|\kappa_1|^2)$, see the components in $|0\rangle$ direction.
	Due to the fact that this $g$ is smaller or equal to $|\kappa_i|^2$, we get the maximal MSEvalue for full separability as
	\begin{align}
		g_{\rm max}=\max\{|\kappa_0|^2,|\kappa_1|^2\}.
	\end{align}

\subsection{Energy Witnesses}
	Let us consider the Hamilton operator $\hat H$ of a system of $N$ harmonic oscillators.
	\begin{align}
		\hat H=-\frac{\hbar^2}{2}\nabla^{\rm T}\boldsymbol M^{-1}\nabla+\frac{1}{2}x^{\rm T}\boldsymbol G x,
	\end{align}
	where we used the position operator vector $\hat x=\left( x_n\right)_{n=1}^N$ and momentum operator vector $\hat p=\frac{\hbar }{\rm i}\nabla=\frac{\hbar }{\rm i}\left(\partial_{x_n}\right)_{n=1}^N$.
	The masses of the individual oscillators are given by
	\begin{align}
		\boldsymbol M={\rm diag}(m_1,\dots,m_N),
	\end{align}
	and the total potential energy reads as $V(x)=x^{\rm T}\boldsymbol G x$, with: $\boldsymbol G\in\mathbb R^{N\times N}$ (real valued); $\boldsymbol G=\boldsymbol G^{\rm T}$ (symmetric); and $\boldsymbol G> 0$ (positive definite).
	The well-known ground state is given by the normalized, Gaussian wave-function
	\begin{align}
		\psi_0(x)=\sqrt[4]{\frac{\det \boldsymbol V}{\pi^N}}\exp\left[-\frac{1}{2}x^{\rm T}\boldsymbol V x \right],\quad
		\text{with }\boldsymbol V=\boldsymbol V^{\rm T}.
	\end{align}
	In order to get $\boldsymbol V$, we need the first and second derivative:
	\begin{align}
		\nabla \psi_0(x)=\left(-\boldsymbol V x\right)\psi_0(x)
		\quad\text{and}\quad
		\nabla\otimes\nabla \psi_0(x)=\left(-\boldsymbol V+\boldsymbol V x\otimes \boldsymbol V x\right)\psi_0(x)
	\end{align}
	The resulting eigenvalue problem is
	\begin{align}
		H\psi_0(x)=&\left[\frac{\hbar^2}{2}{\rm Tr}\!\left(\boldsymbol M^{-1}\boldsymbol V\right)\right]\psi_0(x)
		+x^{\rm T}\left[-\frac{\hbar^2}{2}\boldsymbol V\boldsymbol M^{-1}\boldsymbol V+\frac{1}{2}\boldsymbol G\right] x\,\psi_0(x)
		=E_0\psi_0(x),
	\end{align}
	where the here used trace operation, ${\rm Tr}$, is acting on $\mathbb R^{N\times N}$ matrices.
	The eigenvalue equation is solved by the ground state $\psi_0(x)$ for
	\begin{align}
		0=-\frac{\hbar^2}{2}\boldsymbol V\boldsymbol M^{-1}\boldsymbol V+\frac{1}{2}\boldsymbol G
		\quad\Leftrightarrow\quad
		\boldsymbol V=\frac{1}{\hbar}\boldsymbol M^{1/2}\left[\boldsymbol M^{-1/2}\boldsymbol G\boldsymbol M^{-1/2}\right]^{\frac{1}{2}}\boldsymbol M^{1/2}.
	\end{align}
	The minimal eigenvalue is
	\begin{align}\label{Eq:EigenValue}
		E_0=\frac{\hbar}{2}\,{\rm Tr}\!\left(\left[\boldsymbol M^{-1/2}\boldsymbol G\boldsymbol M^{-1/2}\right]^{\frac{1}{2}}\right).
	\end{align}

	As an intermediate step, it is useful to show that a minimal energy in our case is always obtained for states with zero mean values for $\hat x$ and $\hat p$.
	We have an energy given by
	\begin{align}
		\langle \hat H\rangle=&{\rm Tr}\!\left(\boldsymbol M^{-1} \langle -\frac{\hbar^2}{2} \nabla\,\nabla^{\rm T}\rangle\right)+\frac{1}{2}{\rm Tr}\!\left(\boldsymbol G \langle x\,x^{\rm T}\rangle\right)\\
		=&{\rm Tr}\!\left(\frac{1}{2}\boldsymbol M^{-1}\left[ \langle(\Delta \hat p)(\Delta \hat p)^{\rm T}\rangle+\langle \hat p\rangle\langle \hat p\rangle^{\rm T} \right]\right)
		+\frac{1}{2}{\rm Tr}\!\left(\boldsymbol G\left[ \langle(\Delta \hat x)(\Delta \hat x)^{\rm T}\rangle+\langle \hat x\rangle\langle \hat x\rangle^{\rm T} \right]\right)\\
		=&\frac{1}{2}\langle \hat p\rangle^{\rm T} \boldsymbol M^{-1}\langle \hat p\rangle+\frac{1}{2}\langle \hat x\rangle^{\rm T} \boldsymbol G\langle \hat x\rangle
		+\frac{1}{2}{\rm Tr}\!\left[\boldsymbol M^{-1}\,{\rm Covar}(\hat p,\hat p^{\rm T})\right]+\frac{1}{2}{\rm Tr}\!\left[\boldsymbol G\,{\rm Covar}(\hat x,\hat x^{\rm T})\right].
	\end{align}
	A local displacement operation gives the smallest value for the case $\langle \hat p\rangle=0=\langle \hat x\rangle$, when considering the same covariance matrices.

	Now, we may study an arbitrary decomposition of $\mathcal I=\{1,\ldots,N\}$:
	\begin{align}
		(\mathcal I_j)_{j=1}^K:\quad
		\mathcal I_j\subset \mathcal I,\quad
		\mathcal I_j\neq \emptyset,\quad
		\mathcal I_j\cap\mathcal I_{j'}=\emptyset \,\,(\text{for }j\neq j'),\quad
		\bigcup_{j=1}^K \mathcal I_j=\mathcal I.
	\end{align}
	We define the position and momentum operator for the corresponding subspace as $\hat x_{(j)}=(x_n)_{n\in\mathcal I_j}$ and $\hat p_{(j)}=\frac{\hbar}{\rm i}(\partial_{x_n})_{n\in\mathcal I_j}$, respectively.
	The block-matrices $\boldsymbol M_{(j,j')}$ and $\boldsymbol G_{(j,j')}$ are similarily obtained from $\boldsymbol M$ and $\boldsymbol G$, by ignoring all rows or columns which are not in $\mathcal I_j$ or $\mathcal I_{j'}$.
	Note that $\boldsymbol M_{(j,j')}=0$ for $j\neq j'$.
	A product wave function of this decomposition $(\mathcal I_j)_{j=1}^K$ reads in position representation as
	\begin{align}
		\psi(x)=\prod_{j=1}^K\psi_{(j)}\left(x_{(j)}\right).
	\end{align}
	Hence, we may write the energy operator as
	\begin{align}
		\hat H=\frac{1}{2}\sum_{j=1}^K \hat p^{\rm T}_{(j)}\boldsymbol M_{(j,j)}^{-1} \hat p_{(j)}+\frac{1}{2}\sum_{j,j'=1}^K \hat x_{(j)}^{\rm T}\boldsymbol G_{(j,j')} \hat x_{(j')}.
	\end{align}
	Now the MSEvalue equations read for a particular choice of $J=1,\dots,K$ as
	\begin{align}
		\hat H_{\psi_{(1)},\ldots,\psi_{(J-1)},\psi_{(J+1)},\ldots,\psi_{(N)}}\psi_{(J)}\left(x_{(J)}\right)=E_0\left[(\mathcal I_j)_{j=1}^K\right]\psi_{(J)}\left(x_{(J)}\right),
	\end{align}
	together with the partially reduced operator
	\begin{align}
		\nonumber \hat H_{\psi_{(1)},\ldots,\psi_{(J-1)},\psi_{(J+1)},\ldots,\psi_{(N)}}=&
		\frac{1}{2}\hat p_{(J)}^{\rm T}\boldsymbol M_{(J,J)}^{-1}\hat p_{(J)}+\frac{1}{2}\sum_{j\neq J}\langle \hat p^{\rm T}_{(j)}\boldsymbol M_{(j,j)}^{-1} \hat p_{(j)}\rangle\hat 1_{(J)}\\
		&+\frac{1}{2}\hat x_{(J)}^{\rm T}\boldsymbol G_{(J,J)} \hat x_{(J)}+\frac{1}{2}\sum_{j\neq J,j'\neq J} \langle \hat x_{(j)}^{\rm T}\boldsymbol G_{(j,j')} \hat x_{(j')}\rangle\hat 1_{(J)},
	\end{align}
	where we used $\langle \hat p\rangle=0=\langle \hat x\rangle$.
	Now the MSEvalue equation for the partition $\mathcal I_J$ is an ordinary eigenvalue problem of a quadratic (reduced) Hamiltonian:
		\begin{align}
			\nonumber&\left[-\frac{\hbar^2}{2}\nabla_{(J)}^{\rm T}\boldsymbol M_{(J,J)}^{-1}\nabla_{(J)}+\frac{1}{2}x_{(J)}^{\rm T}\boldsymbol G_{(J,J)} x_{(J)}\right]\psi_{(J)}\left(x_{(J)}\right)\\
			&+\left[\frac{1}{2}\sum_{j\neq J}\langle \hat p^{\rm T}_{(j)}\boldsymbol M_{(j,j)}^{-1} \hat p_{(j)}\rangle+ \frac{1}{2}\sum_{j\neq J} \langle \hat x_{(j)}^{\rm T}\boldsymbol G_{(j,j)} \hat x_{(j')}\rangle\right]\psi_{(J)}\left(x_{(J)}\right)
			=E_0\left[(\mathcal I_j)_{j=1}^K\right]\psi_{(J)}\left(x_{(J)}\right).
		\end{align}
	From our initial solution of the ground state of the full Hamiltonian, we obtain the ground state of the reduced Hamiltonian being the MSEvector in position representation,
	\begin{align}
		\psi_{(J)}(x_{(J)})=\sqrt[4]{\frac{\det \boldsymbol V_{\!J}}{\pi^{N_{\!J}}}}\exp\left[-\frac{1}{2}x_{(J)}^{\rm T}\boldsymbol V_{\!J} x_{(J)} \right],
		\quad\text{for }
		\boldsymbol V_{\!J}=\frac{1}{\hbar}\boldsymbol M_{(J,J)}^{1/2}\left[\boldsymbol M_{(J,J)}^{-1/2}\boldsymbol G_{(J,J)}\boldsymbol M_{(J,J)}^{-1/2}\right]^{\frac{1}{2}}\boldsymbol M_{(J,J)}^{1/2},
	\end{align}
	with the cardinality $N_{\!J}$ of $\mathcal I_J$, $N_{\!J}=|\mathcal I_{J}|$.
	This yields the corresponding minimal MSEvalue as
	\begin{align}
		E_0\left[(\mathcal I_j)_{j=1}^K\right]=\sum_{j=1}^K E_0\left[\mathcal I_j\right]=\sum_{j=1}^K\frac{\hbar}{2}\,{\rm Tr}\!\left(\left[\boldsymbol M_{(j,j)}^{-1/2}\boldsymbol G_{(j,j)}\boldsymbol M_{(j,j)}^{-1/2}\right]^{\frac{1}{2}}\right).
	\end{align}
	Let us note, that we get for $K=1$ $E_0\left[(\mathcal I)\right]=E_0$, cf. Eq.~\eqref{Eq:EigenValue}.

	The example in the paper has the following properties:
	\begin{align}
		\boldsymbol M=&m\,{\rm diag}(\boldsymbol 1_1,\dots,\boldsymbol 1_N),\\
		\boldsymbol G=&\left[m\omega^2+N\gamma\right]{\rm diag}(\boldsymbol 1_1,\dots,\boldsymbol 1_N)-\left[N\gamma\right]\vec n\,\vec n^{\rm T}\\
		\vec n=&\frac{1}{\sqrt{N}}\left(\boldsymbol 1_i\right)_{i=1}^{N}\\
		\boldsymbol 1_i=&\vec e_{i,x}\,\vec e_{i,x}^{\,\rm T}+\vec e_{i,y}\,\vec e_{i,y}^{\,\rm T}+\vec e_{i,z}\,\vec e_{i,z}^{\,\rm T}.
	\end{align}
	Note that each component is a 3 dimensional matrix itself and the potential energy is $V(\vec r_1,\dots,\vec r_N)=\frac{1}{2}\left[(\vec r_j)_{j=1}^N\right]^{\rm T}\boldsymbol G\left[(\vec r_j)_{j=1}^N\right]$.
	For a given $\mathcal I_j$, we have
	\begin{align}
		\vec n_{(j)}=&\frac{1}{\sqrt{|\mathcal I_j|}}\left(\boldsymbol 1_i\right)_{i\in\mathcal I_j}\\
		\nonumber E_0[\mathcal I_j]=&
		\frac{\hbar}{2}\,{\rm Tr}\!\left(\left[\boldsymbol M_{(j,j)}^{-1/2}\boldsymbol G_{(j,j)}\boldsymbol M_{(j,j)}^{-1/2}\right]^{\frac{1}{2}}\right)\\
		\nonumber =&\frac{\hbar}{2\sqrt{m}}\,{\rm Tr}\!\left(\left(
		\left[m\omega^2+N\gamma\right]\left[{\rm diag}(\boldsymbol 1_1,\dots,\boldsymbol 1_N)-\vec n_{(j)}\,\vec n_{(j)}^{\rm T}\right]+\left[m\omega^2+\gamma(N-|\mathcal I_j|)\right]\vec n_{(j)}\,\vec n_{(j)}^{\rm T}
		\right)^{\frac{1}{2}}\right)\\
		\nonumber =&\frac{\hbar}{2\sqrt{m}}\,{\rm Tr}\!\left(
		\left[m\omega^2+N\gamma\right]^{\frac{1}{2}}\left[{\rm diag}(\boldsymbol 1_i)_{i\in\mathcal I_j}-\vec n_{(j)}\,\vec n_{(j)}^{\rm T}\right]
		+\left[m\omega^2+\gamma(N-|\mathcal I_j|)\right]^{\frac{1}{2}}\vec n_{(j)}\,\vec n_{(j)}^{\rm T}\right)\\
		=&\frac{\hbar\omega}{2}\left(\left[1+\frac{\gamma N}{m\omega^2}\right]^{\frac{1}{2}}\left[3|\mathcal I_j|-3\right]
		+\left[1+\frac{\gamma (N-|\mathcal I_j|)}{m\omega^2}\right]^{\frac{1}{2}}3\right),
	\end{align}
	where the second line of $E_0[\mathcal I_j]$ represents the spectral decomposition of $\boldsymbol G_{(j,j)}$.

\end{widetext}	

\end{document}